\documentclass[aps,pra,twocolumn,showpacs,preprintnumbers,amsmath,amssymb]{revtex4-1}
\usepackage{latexsym}
\usepackage{graphicx}
\usepackage{graphics}
\usepackage[T1]{fontenc}
\usepackage{multirow}

\newcommand{\ket}[1]{\mbox{$ | #1 \rangle $}}

\newcommand{\module}[1]{\left|#1\right|}

\newcommand{\Sectionname}[1]{\textsc{Sec.~}#1}

\begin{document}

\author{Florian Kaiser}
\author{Amandine Issautier}
\author{Lutfi A. Ngah}
\author{Olivier Alibart}
\author{Anthony Martin}
\author{S\'ebastien Tanzilli}\email{sebastien.tanzilli@unice.fr}
\affiliation{Laboratoire de Physique de la Matière Condensée, CNRS UMR 7336, Université Nice -- Sophia Antipolis, Parc Valrose, 06108 Nice Cedex 2, France}

\title{A versatile source of polarisation entangled photons for quantum network applications}

\begin{abstract}
We report a versatile and practical approach for generating high-quality polarization entanglement in a fully guided-wave fashion. Our setup relies on a high-brilliance type-0 waveguide generator producing paired photon at a telecom wavelength associated with an advanced energy-time to polarisation transcriber. The latter is capable of creating any pure polarization entangled state, and allows manipulating single photon bandwidths that can be chosen at will over five orders of magnitude, ranging from tens of MHz to several THz.
We achieve excellent entanglement fidelities for particular spectral bandwidths, \textit{i.e.} 25\,MHz, 540\,MHz and 100\,GHz, proving the relevance of our approach. Our scheme stands as an ideal candidate for a wide range of network applications, ranging from dense division multiplexing quantum key distribution to heralded optical quantum memories and repeaters.
\end{abstract}

\pacs{03.65.Ud, 03.67.Bg, 03.67.Hk, 03.67.Mn, 42.50.Dv, 42.65.Lm, 42.65.Wi}
\keywords{Integrated optics; Nonlinear optics; Quantum Optics; Entanglement; Quantum Information \& Communication}

\maketitle

\section{Introduction}

Entanglement served historically as an essential resource for testing the foundations of quantum physics, such as non-locality~\cite{EPR_1935} via the violation of the Bell inequalities~\cite{Bell_EPR_1964,BCHSH_1969,aspect_experimental_1982}. More recently, entanglement has been employed as a key ingredient in extended versions of Bohr's complementarity notion~\cite{Peruzzo_QDC_2012,Kaiser_QDC_2012}.
Today, quantum information science (QIS) exploits entanglement for enhanced communication and processing protocols~\cite{tittel_photonic_2001}. On one hand, quantum key distribution (QKD), already commercialized, provides a unique means to establish private ciphers between distant partners~\cite{Scarani_QKD_2009}. On the other hand, various entanglement-enabled network tasks, such as quantum relays~\cite{Aboussouan_dipps_2010}, memories~\cite{lvovsky_optical_2009}, and repeaters~\cite{Sangouard_DLCZRMP_2011}, are extensively studied in research laboratories.
State-of-the-art manipulation of entanglement is achieved using disparate experimental techniques such as atom traps~\cite{Lettner_remote_2011}, Josephson junctions~\cite{Kubo_supercond_2010}, and quantum integrated photonics~\cite{Tanzilli_genesis_2012}. System versatility and compatibility stand now as corner stones for pushing QIS one step further.
In the context of photonic entanglement sources~\cite{Kuklewicz_2006,Halder_2008,Piro_SPSA_2009,Pomarico_2009,Yan_AtomsEntanglement_2011,Kaiser_typeII_2012}, we report a remarkably versatile solution, capable of creating any pure polarisation entangled state encoded on telecom wavelength photons.
Exploiting an advanced energy-time to polarisation observable \textit{transcriber} associated with a wavelength tunable high-brilliance photon-pair generator, we demonstrate excellent entanglement fidelities for spectral bandwidths that can be chosen over five orders of magnitude. Our scheme stands an ideal candidate for a wide range of network applications, ranging from dense division multiplexing QKD~\cite{Yoshino_DWDMQKD_2012} to heralded optical quantum memories~\cite{lvovsky_optical_2009}.

\begin{figure*}
\centering
\includegraphics[width=1.8\columnwidth]{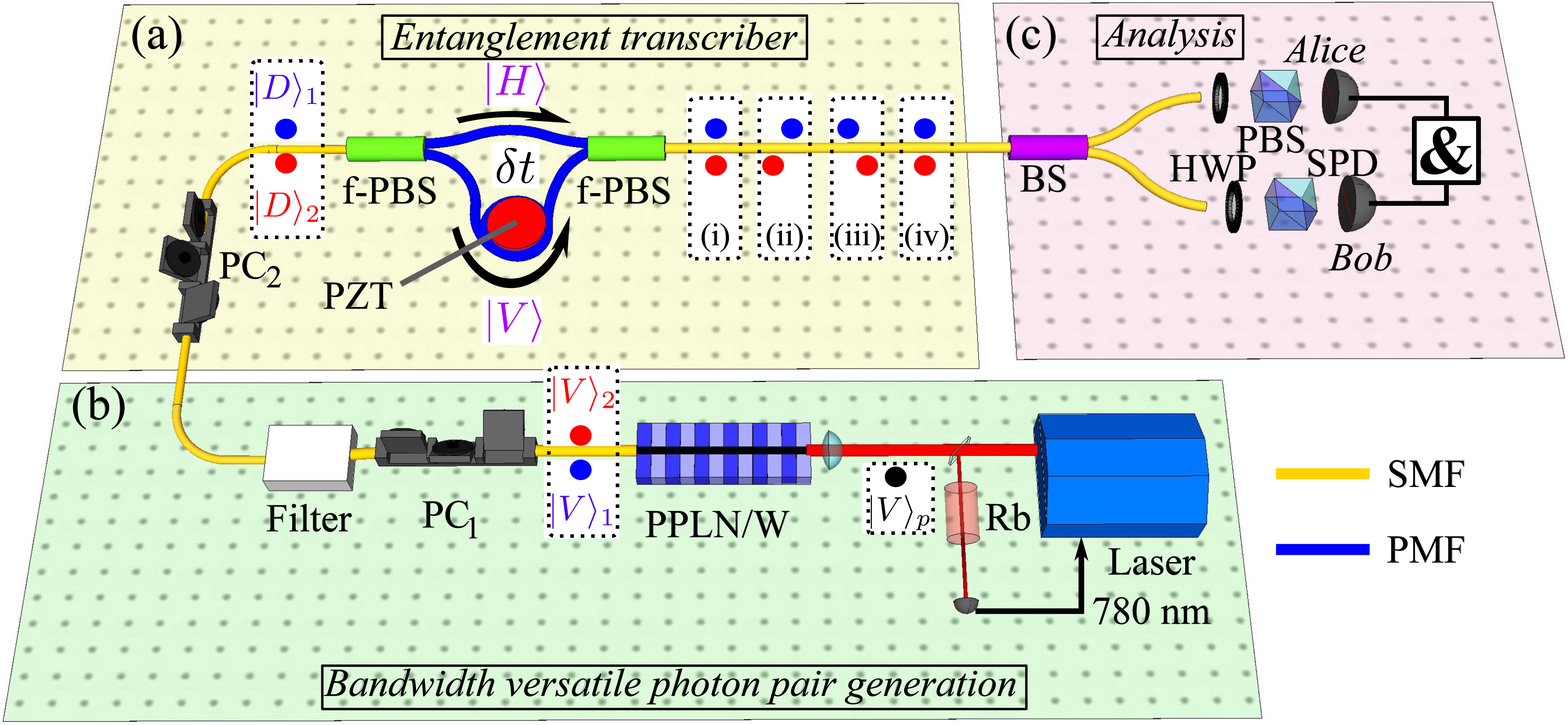}
\caption{\label{fig1}Schematic of the experimental setup.
(a) Fibre optical transcriber apparatus. Sending the diagonally polarised bi-photon state $\ket{D}_1 \otimes \ket{D}_2 = \frac{1}{\sqrt{2}}(\ket H_1 + \ket V_1) \otimes \frac{1}{\sqrt{2}}(\ket H_2 + \ket V_2)$, leads to four possible outputs: (i) $\ket{V,l}_1\ket{V,l}_2$, (ii) $\ket{V,l}_1\ket{H,e}_2$, (iii) $\ket{H,e}_1\ket{V,l}_2$ and (iv) $\ket{H,e}_1\ket{H,e}_2$, where $e$ and $l$ denote 'early' and 'late' temporal modes, respectively. Temporal post-selection of cases (i) and (iv) generates the desired polarisation entangled state $\ket{\Phi(\phi)}$.
(b) High-efficiency photon-pair generator. A 780\,nm laser pumps a PPLN/W to generate paired-photons at the degenerate wavelength of 1560\,nm. After passing through a polarisation controller (PC$_1$), the photons are conveniently filtered to bandwidths ranging from 25\,MHz to 100\,GHz, using standard telecom techniques.
(c) Standard polarisation state analysers. Alice and Bob comprise each a half-wave plate (HWP), a polarising beam-splitter (PBS), and a single-photon detector (SPD). \&: AND-gate that connects the two SPDs for recording coincidence counts between the two users. PZT: piezoelectric transducer fibre stretcher.}
\end{figure*}

\section{Experimental setup and related capabilities}

As in classical communication, low-loss optical fibres and high-performance components based on guided-wave technologies enable exploiting the telecom C-band wavelengths (1530-1565\,nm) for generating and distributing photonic entanglement. Among the most widely used observables are time-bin and polarisation~\cite{tittel_photonic_2001}, the latter being undoubtedly the easiest to analyse.
Depending on the application, many schemes have been developed to generate photonic polarisation entanglement. Usually, the photon pairs are generated using the nonlinear process of spontaneous parametric down-conversion (SPDC) in either bulk~\cite{Piro_SPSA_2009} or waveguide configurations~\cite{Kaiser_typeII_2012}, possibly surrounded by an optical cavity for reducing the generated bandwidth~\cite{Kuklewicz_2006,Pomarico_2009}. Among other pertinent generators, we find dispersion-shifted fibres~\cite{Fulconis_PCF_2007,Medic_Fiber_2010}, quantum dots~\cite{dousse_Qdotultrabright_2010}, and alkaline cold atomic ensembles~\cite{Dudin_TelecomColdAtom_2010,Yan_AtomsEntanglement_2011}. However up to now, this has been made at the price of either low brightness~\cite{Kuklewicz_2006,Piro_SPSA_2009,Kaiser_typeII_2012}, limited entanglement fidelities~\cite{Fulconis_PCF_2007,dousse_Qdotultrabright_2010}, or complex equipment~\cite{Medic_Fiber_2010,Dudin_TelecomColdAtom_2010,Yan_AtomsEntanglement_2011}. Moreover, none of these realisations has demonstrated versatility, since every new application requires a new adapted source.

We overcome these problems with the experimental setup of \figurename{~\ref{fig1}}. It consists of two main stages, namely an advanced fibre optical energy-time to polarisation transcriber apparatus (a) and a wavelength tunable, high-brilliance, periodically poled lithium niobate waveguide (PPLN/W) generator (b). The transcriber is used to introduce a delay $\delta t$ between horizontal ($\ket H$) and vertical ($\ket V$) polarisation modes. Sending the diagonally ($\ket D$) polarised bi-photon state $\ket{\psi}_{\rm in} = \ket D_1 \otimes \ket D_2 = \frac{1}{\sqrt{2}}(\ket H_1 + \ket V_1) \otimes \frac{1}{\sqrt{2}}(\ket H_2 + \ket V_2)$, where the subscripts $\{1,2\}$ denote the two incoming photons that may differ in wavelength and/or in emission time, results in four temporal output contributions. As shown in \figurename{~\ref{fig1}}(a), post-selecting temporally the bi-photon zero-delay contributions (i) and (iv) reduces $\ket{\psi}_{\rm in}$ to the maximally entangled state $\ket{\Phi(\phi)} = \frac{1}{\sqrt{2}} \left( \ket{H}_1 \ket{H}_2 + e^{i\phi} \ket{V}_1\ket{V}_2 \right)$, where $\phi$ is twice the relative phase between vertical and horizontal polarisation modes.
A detailed analysis of the transcriber operation, and its generalization for creating non-maximally entangled states, are given in \Sectionname{~\ref{Sec_Ann_Trans}}.
Our transcriber is made of short and long polarisation maintaining fibres (PMF) connecting two fibre polarising beam-splitters (f-PBS).
The temporal mode separation, given by the path length difference between the two arms, is set to $\delta t=76$\,ns ($\delta L = 18$\,m). This allows manipulating and further entangling ultra-long coherence time ($\tau_{c}^{{\rm phot}}$) single-photons, \textit{i.e.} up to 23\,ns ($\leftrightarrow$ 19\,MHz).
Compared to previous transcriber-like realisations~\cite{Ribordy_QKD_2000,Sanaka_BDL_2002}, this represents more than 50 times improvement, enabled by an active and high-speed phase stabilization scheme. It allows accurate and on-demand phase control of the created entangled states over long time scales, as discussed in \Sectionname{~\ref{Sec_Ann_Phase}}.

\section{The integrated optics waveguide generator and related characterization}

\begin{figure}[h!]
\centering
\includegraphics[width=1\columnwidth]{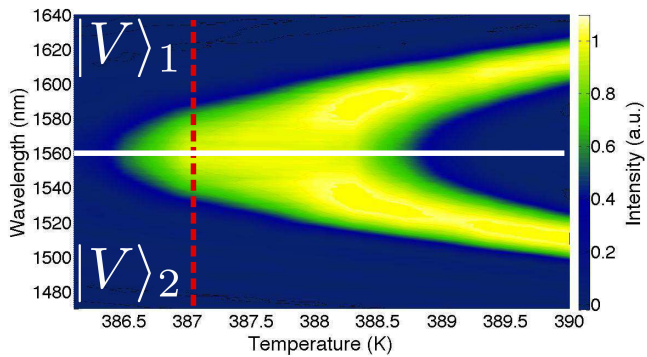}
\caption{\label{fig2}Wavelength tunability of the photon-pair generator measured as a function of the temperature. 3\,K temperature tuning results in a wavelength tunability of more than 100\,nm. The horizontal line represents the degeneracy point where the two photons have the same wavelength of 1560\,nm, and corresponds to the working operation of our experiment. At this particular point, obtained for a temperature of 387\,K (vertical dashed line), the emission bandwidth is of 4\,THz.}
\end{figure}

We now connect the transcriber to a 4.5\,cm long PPLN/W photon-pair generator exploiting the so-called type-0 SPDC process.
As shown in \figurename{~\ref{fig1}}(b), the crystal is pumped by a vertically polarised 780\,nm continuous wave laser for creating vertically polarised paired photons $\ket V_1 \ket V_2$ at an average flux as high as $\sim 10^{10}$ pairs per second and per mW of pump power. Due to the spontaneous character of the down-conversion process, the coherence time of the pairs corresponds to that of the laser. To fulfil the transcriber requirement $\tau_c^{\rm pair} \gg \delta t$, the laser is frequency stabilized against a hyperfine transition in the $D_2$ line of rubidium, giving a pair coherence time $\tau_c^{\rm pair} = 3\,\rm \mu$s. As shown in \figurename{~\ref{fig2}}, wavelength degeneracy, \textit{i.e.} 1560\,nm, is obtained at the temperature of 387\,K, and the associated spectral bandwidth is of about 4\,THz ($\leftrightarrow$ 32\,nm), covering the full telecom C-band of wavelengths.

In addition, the emission wavelength can be tuned over more than 100\,nm, with respect to the SPDC energy conservation, by adapting the phase-matching condition via temperature control.
After the PPLN/W, the pairs are collected using a single mode fibre and sent to the filtering stage.
In this realisation, we use three exemplary filters: a standard 100\,GHz dense wavelength division multiplexer (DWDM) adapted to standard telecom networks~\cite{Yoshino_DWDMQKD_2012}, a 540\,MHz phase-shifted fibre Bragg grating (PS-FBG) compatible with broadband spectral acceptance quantum memories~\cite{reim_towards_2010,Saglamyurek_BroadbandWQM_2011} and continuous-wave quantum relays~\cite{Halder_2008}, and a 25\,MHz PS-FBG matching much narrower spectral acceptance quantum memories~\cite{Tanji_HQM_2009,Clausen_QSPECrystal_2011}. This results in a source bandwidth versatility covering more than five orders of magnitude by simple filter adaptation. More details on the PS-FBG filters are given in \Sectionname{~\ref{Sec_Ann_FBG}}.
After the filter, polarisation entanglement is created by the transcriber, in front of which a polarisation controller (PC$_2$) is used to rotate the photons' polarisation to the diagonal state $\ket{\psi}_{\rm in} = \ket D_1 \ket D_2$, as described above. This condition is necessary for generating the maximally entangled state $\ket{\Phi(\phi)}$, and therefore violating the Bell inequalities~\cite{Bell_EPR_1964,BCHSH_1969} with optimal visibilities~\cite{Martin_Analysis_2012}.

\section{Coincidence histogram for three different bandwidths}

\begin{figure}[h!]
\centering
\includegraphics[width=1\columnwidth]{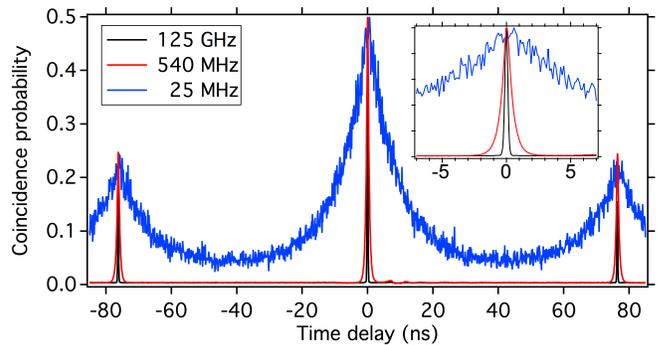}
\caption{Second order intensity correlation measurement after the transcriber apparatus. The central coincidence peak contains the contributions to the entangled state $\ket H_1 \ket H_2$ and $\ket V_1 \ket V_2$. For all the utilized photon-pair bandwidths, \textit{i.e.} 100\,GHz (black curve), 540\,MHz (red), and 25\,MHz (blue), the central peak is well separated from the side peaks that are associated with paired photons in a separable state. The current peak separation of 76\,ns is sufficient to distinguish central and side peaks down to spectral bandwidths of 19\,MHz. The right inset is a zoom on the central peak.\label{fig3}}
\end{figure}

To characterize the suitability of the transcriber for handling narrowband photons, we first measure the time-dependent two-photon correlation function for the three filters mentioned above. A 50/50 fibre beam-splitter (BS) is used to separate and distribute the photons to Alice and Bob, each employing a single-photon detector (SPD) connected to a coincidence measurement apparatus (\&). More details on these detectors are provided in \Sectionname{~\ref{Sec_Ann_SPD}}. As shown in \figurename{~\ref{fig3}}, three well separated coincidence peaks are obtained for each filter.

The two outer peaks at $\delta t = \pm$76\,ns correspond to cross polarisation contributions which have been split up temporally by the transcriber, while the central peak contains the $\ket H_1 \ket H_2$ and $\ket V_1 \ket V_2$ contributions to the desired entangled state. At this stage, the central-to-side peak ratio of 2 indicates that all the contributions have the same probability amplitudes and that maximally entangled states can be post-selected out of the central peak. For the 100\,GHz bandwidth, a $\sim$4\,ps photon coherence time is expected, such that the 230\,ps peak width is mainly given by the convolution of the detectors' timing jitters. With the two narrowband filters in place, the single photons' coherence time is increased. This leads to a peak broadening of 800$\pm$20\,ps and 15.6$\pm$0.7\,ns, which is in good agreement with the specified bandwidths of the 540 and 25\,MHz filters, respectively.

\section{Entanglement quality three different bandwidths}

We now examine the quality of the produced entanglement at the output of the full setup of \figurename{~\ref{fig1}}. For state analysis (c), we use a standard Bell inequality type setup, where Alice and Bob employ each a half-wave plate (HWP), a polarising beam-splitter (PBS), and an SPD. We also consider only photon-pair events within the full-width at half-maximum region of the central coincidence peaks of \figurename{~\ref{fig3}} for post-selecting the maximally entangled state $\ket{\Phi(\phi)}$.
Violating the Bell inequalities requires maintaining the coherence of the state, \textit{i.e.} in our case stabilizing the phase $\phi$, over the full measurement time~\cite{Martin_Analysis_2012}. We also stress that, for our extremely long temporal mode separation, the transcriber is required to be an advanced system, meaning that a high-speed ($>$1\,kHz) and high-resolution ($\Delta \phi < \frac{\pi}{100}$) phase stabilization scheme is employed. This is detailed and demonstrated to be remarkably suitable to that purpose in \Sectionname{~\ref{Sec_Ann_Phase}}.

\begin{figure}[h!]
\centering
\includegraphics[width=1\columnwidth]{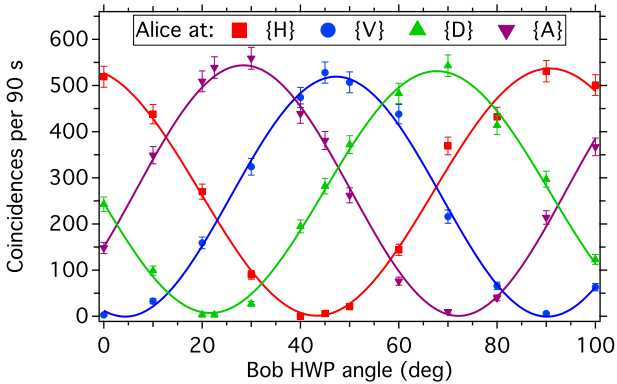}
\caption{Violation of the Bell inequalities for the state $\ket{\Phi^-}$ with the 25\,MHz filter. For the four settings on Alice's side, excellent raw visibilities are demonstrated when rotating continuoulsly Bob's HWP. Lines in the graph represent sinusoidal fits.\label{fig4}}
\end{figure}

In the following, we analyse the state $\ket{\Phi^-}$, obtained when $\phi$ is set to $\pi$ in the transcriber. In this case, the coincidence rate is measured for four standard and consecutive analysis settings on Alice's side, \textit{i.e.} $\ket{H}$, $\ket{V}$, $\ket{D}$, and $\ket{A}$, while Bob's HWP is continuously rotated.
As an exemplary result, we consider the 25\,MHz filter and take advantage, for this particular measurement, of two super-conducting SPDs for reducing accidental coincidence events (for more details, see \Sectionname{~\ref{Sec_Ann_SPD}}). As shown in \figurename{~\ref{fig4}}, excellent raw visibilities, of 99$\pm$3\%, are obtained for all the four orientations at Alice's.
The corresponding Bell parameter is calculated to be $S_{raw}=2.82\pm0.02$, which leads to a violation of the Bell inequalities by more than 40 standard deviations~\cite{Bell_EPR_1964,BCHSH_1969}. For the two other spectral bandwidths, \textit{i.e.} 100\,GHz and 540\,MHz, similar high visibilities have been measured. All the obtained  entanglement measurement results and other pertinent figures of merit are summarized in \tablename{~\ref{Table_Source_type0}}.

\begin{table}[h!]
\begin{center}
\begin{tabular}{| l || c | c | c |} \hline
Bandwidth [MHz] & 125$\cdot 10^{3\dagger}$ & 540$^\dagger$ & 25$^\ddagger$ \\ \hline
$\tau_{c}^{{\rm phot}\,\ast}$ [ns] & 4.4$\cdot 10^{-3}$ & 0.8 & 15.6 \\ \hline
Pump power$^\star$ [mW] & 0.02 & 0.6 & 7 \\ \hline
$\frac{\bar{n}}{\tau_{c}^{{\rm phot}}}^\divideontimes$ [ns$^{-1}$] & 5$\cdot 10^{-5}$ & 2$\cdot 10^{-3}$ & 2$\cdot 10^{-2}$ \\ \hline
Detected pair rate [s$^{-1}$] & 2000 & 50 & 6 \\ \hline
B [(s$\cdot$mW$\cdot$MHz)$^{-1}$] & 960 & 300 & 380 \\ \hline
$V_{\rm raw}$ (\%) & 99.6\,$\pm1.3$ &  97.1\,$\pm0.9$  & 99\,$\pm$3 \\ \hline
$\mathcal{F}^{\ket{\Phi^-}}_{\rm raw}$ & 0.998 & 0.985 & 0.995 \\ \hline
${S}_{\rm raw}$ & 2.82$\pm$0.01 & 2.80$\pm$0.02 & 2.82$\pm$0.02 \\ \hline
\end{tabular}
\caption{\label{Table_Source_type0}Summary of the entanglement measurements carried out with the source setup of \figurename{~\ref{fig1}}. The main results are given as a function of the considered spectral bandwidth, and without any correction for noise contributions (raw);
$V_{\rm raw}$: two-photon interference pattern visibility obtained in a standard Bell inequality test setup; $\mathcal{F}^{\ket{\Phi^-}}_{\rm raw}$: fidelity to the closest maximally entangled state $\ket{\Phi^-}$; ${S}_{\rm raw}$: corresponding Bell parameter~\cite{Bell_EPR_1964}, calculated as introduced by Clauser, Horne, Shimony, and Holt~\cite{BCHSH_1969}.
$^{\ast}$\,$\tau_{c}^{{\rm phot}}$: single-photon coherence time.
$^\star$\,The pump powers are measured in front of the waveguide generator.
$^\divideontimes$\,$\frac{\bar{n}}{\tau_{c}^{{\rm phot}}}$: Mean number of pairs per coherence time.
$^\dagger$\,For these two configurations, two indium-gallium-arsenide free-running avalanche photodiodes have been employed as single-photon detectors (SPD)~\cite{Hadfield_SPD_2009}.
$^\ddagger$\,For this configuration, two super-conducting SPDs have been employed for reducing accidental coincidence events~\cite{Hadfield_SPD_2009}.
Note that the brightness (B) unit corresponds to the number of pairs of photons coupled in a single-mode fibre, normalized per second, mW of pump power, and per MHz of spectral bandwidth.}
\end{center}
\end{table}

\section{Conclusion}

We have implemented a remarkably versatile experiment towards producing polarisation entangled photons. The versatility concerns both the created state as well as the spectral properties of the photons. In other words, taking advantage of an advanced fibre optical transcriber connected to a highly efficient waveguide generator, we can create any pure polarisation entangled state. Depending on the application, by adapting the phase-matching condition and the filter, the single photon wavelength can be tuned over more than 100\,nm and associated spectral bandwidth chosen over more than five orders of magnitude.
We have also achieved remarkable normalized source brightnesses which stand among the highest ever reported for narrowband entangled photon-pairs~\cite{Kuklewicz_2006,Halder_2008,Piro_SPSA_2009,Pomarico_2009,Yan_AtomsEntanglement_2011,Kaiser_typeII_2012}. Analysing the maximally entangled state $\ket{\Phi^-}$, the Bell inequalities have been violated by more than 40 standard deviations, for three exemplary filters, \textit{i.e.} 100\,GHz, 540\,MHz, and 25\,MHz.

We believe that such a versatile and high-performance realisation represents an ideal candidate for implementing fundamental quantum optics experiments as shown in~\cite{Kaiser_QDC_2012}, as wall as various quantum network scenarios, ranging from dense division multiplexing QKD (100\,GHz to 1\,THz) to heralded optical quantum memories (10\,MHz to 1\,GHz)~\cite{lvovsky_optical_2009}. In the latter framework, our source is already suitable for connection with future ion or atom based quantum storage devices operating directly at a telecom wavelength. With current quantum storage devices operating below 900\,nm, a coherent wavelength adaptation, in and out of the memories, can be addressed by means of non-linear optical frequency conversion~\cite{Tanzilli_Qinterface_2005,Curtz_UpSingle_2010,Dudin_TelecomColdAtom_2010}.

\section*{Acknowledgements}
The authors thank V. D'Auria, A. Kastberg, L. Labont\'{e}, M. P. De Micheli, and D. B. Ostrowsky for their help, as well as ID Quantique, Scontel, AOS GmbH, Teraxion, and OLI (Toptica Photonics) for technical support. Financial support from the CNRS, the ANR ``e-QUANET'' project (grant ANR-09-BLAN-0333-01), the European ICT-2009.8.0 FET open project ``QUANTIP'' (grant 244026), the MESR, the DGA, the Conseil R\'egional PACA, and the MARA, is acknowledged.




\section{Appendix}

\subsection{Fibre optical transcriber for producing maximally and non-maximally entangled states}
\label{Sec_Ann_Trans}

First consider, at the transcriber input, the diagonally polarised bi-photon state $\ket{\psi}_\mathrm{in} = \ket D_1 \otimes \ket D_2 = \frac{1}{\sqrt{2}}(\ket H_1 + \ket V_1) \otimes \frac{1}{\sqrt{2}}(\ket H_2 + \ket V_2)$, where $\ket{D}_i$ represents the diagonal polarisation state, and the subscripts $\{1,2\}$ the two photons that could differ in wavelength and/or in emission time.
After the transcriber, the state reads $\ket{\psi}_\mathrm{out} = \frac{1}{\sqrt{2}} \left(\ket{H,e}_1 + e^{i\phi_1}\ket{V,l}_1\right)\otimes \frac{1}{\sqrt{2}} \left(\ket{H,e}_2 +e^{i\phi_2} \ket{V,l}_2\right)$, where $e$ and $l$ denote early and late temporal modes, respectively, separated by $\delta t$ with a relative phase difference $\phi_i$, $i\in \{1,2\}$. From the experimental side, separating the paired photons at a beam-splitter and recording coincidence counts between two detectors leads to the histogram of \figurename{~\ref{fig3}}. The operation principle of the transcriber is similar to that of energy-time entanglement~\cite{tittel_photonic_2001}. On one hand, $\delta t$ is required to be much greater than the coherence time of the single photons ($\tau_c^{\rm phot}$) for preventing temporal overlap between pairs projected onto parallel ($\ket{H,e}_1 \ket{H,e}_2$ or $\ket{V,l}_1 \ket{V,l}_2$) and orthogonal ($\ket{H,e}_1 \ket{V,l}_2$ and $\ket{V,l}_1 \ket{H,e}_2$) states. On the other hand, the coherence time of the incident pairs ($\tau_c^{\rm pair}$) has to be substantially greater than $\delta t$ for allowing interference between early and late two-photon contributions in the central peak of \figurename{~\ref{fig3}}. Post-selecting these central peak events reduces $\ket{\psi}_\mathrm{out}$ to the maximally entangled state $\ket{\Phi(\phi)} = \frac{1}{\sqrt{2}} \left( \ket{H}_1 \ket{H}_2 + e^{i\phi} \ket{V}_1\ket{V}_2 \right)$, where $\phi=\phi_1+\phi_2$.\\
This scheme can be generalized when considering, at the transcriber input, a two-qbit product state of the form $\ket{\psi}_\mathrm{in} = \left(\alpha_1 \ket H_1 + \beta_1 \ket V_1 \right) \otimes \left( \alpha_2 \ket H_2 + \beta_2 \ket V_2 \right)$, where $\module{\alpha_i}^2 + \module{\beta_i}^2 = 1$. In this case, appropriate post-selection projects the state onto $\ket{\psi}_\mathrm{out}  = \alpha \ket{H}_1 \ket{H}_2 + \beta e^{i\phi} \ket{V}_1\ket{V}_2$, where $\alpha= \frac{\alpha_1 \alpha_2}{\sqrt{\module{\alpha_1 \alpha_2}^2 +\module{\beta_1\beta_2}^2}}$ and $\beta = \frac{\beta_1\beta_2}{\sqrt{\module{\alpha_1 \alpha_2}^2 +\module{\beta_1\beta_2}^2}}$. Consequently, this scheme permits creating any pure bi-photon polarisation entangled state, expressed as superpositions of $\ket{\Phi^+}$ and $\ket{\Phi^-}$ Bell states. Note that in our case, the three parameters $\alpha$, $\beta$, and $\phi$ are accessible experimentally, namely by controlling the input state and fine tuning of the phase set in the transcriber. Furthermore, rotating the polarisation of one photon by $\pi/2$ (using an additional half-wave plate) after the transcriber, and suitably choosing $\phi$ between $0$ and $\pi$, allows creating superpositions of $\ket{\Psi^+}$ and $\ket{\Psi^-}$ Bell states.

\subsection{Phase stabilization of the transcriber, and manipulation of the entangled state}
\label{Sec_Ann_Phase}

\begin{figure}
\centering
\includegraphics[width=1\columnwidth]{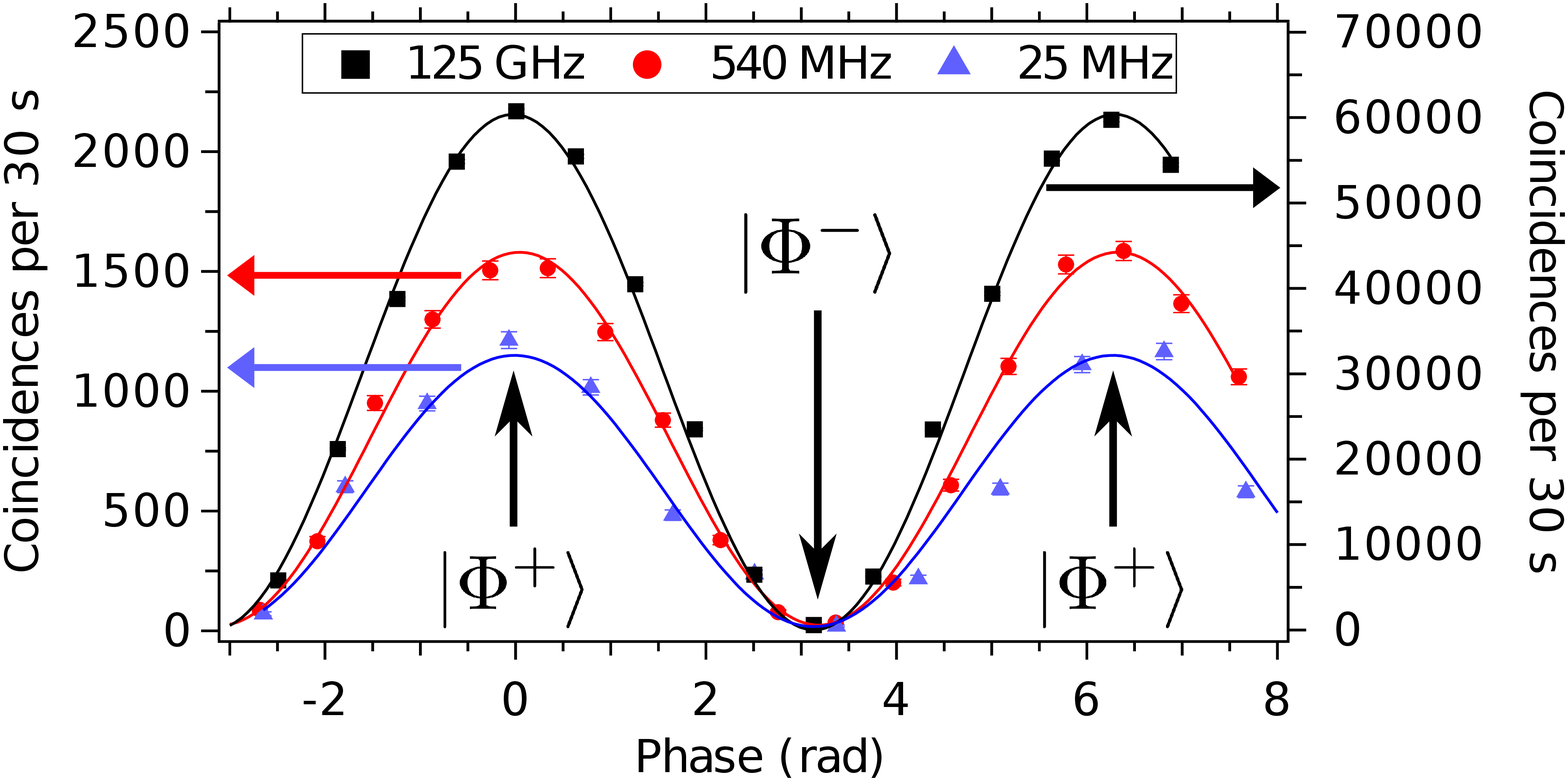}
\caption{Coincidence rates as a function the phase $\phi$ set in the transcriber. The black, red, and blue curve represents the recorded data for the 100\,GHz, 540\,MHz, and 25\,MHz, respectively. For the three curves, accidental coincidence events due to the dark-counts in the detectors have been subtracted, leading to the direct observation of net visibilities. 
The maxima and minima of the coincidence rates are associated with the Bell states $\ket{\Phi^+}$ and $\ket{\Phi^-}$, respectively. Lines in the graph represent sinusoidal fits.\label{fig5}}
\end{figure}

Violating the Bell inequalities requires the coherence, \textit{i.e.} the phase relation $\phi$ between the two contributions to the entangled state, to be stable during the full measurement~\cite{Martin_Analysis_2012}. Phase fluctuations mainly come from temperature fluctuations in the long arm of the transcriber. For 18\,m path length difference, we have $\Delta \phi / \Delta T \approx 10^3\,\rm rad/K$. Stabilizing the phase via temperature control only would require sub-mK temperature stability, which is technically challenging, especially for long-term measurements. On the contrary to former transcriber-like realisations~\cite{Ribordy_QKD_2000,Sanaka_BDL_2002} that relied, at most, on temperature stabilization, we actively stabilize the optical length of the retardation line using a 50\,kHz feedback loop system. It comprises a piezoelectric transducer (PZT) fibre stretcher in the long arm to compensate for drifts, and a telecom reference laser to constantly monitor the phase. The laser is actively frequency stabilized with respect to the frequency of the pump laser using a frequency doubling stage followed by a transfer cavity locking scheme. A fraction of this laser light is sent to the transcriber in the backward direction compared to that of the paired photons. With such a stabilization system, the phase in the transcriber can be set on demand to any desired value, and further reconfigured in a very fast ($>$\,1\,kHz) and accurate ($\Delta \phi < \frac{\pi}{100}$) manner.\\
\indent
To demonstrate our capability to both control and accurately tune the phase $\phi$ set in the transcriber, Alice and Bob fix their respective half-wave plate (HWP, see \figurename{~\ref{fig1}}) at $22.5^{\circ}$ to project the entangled state in the phase sensitive diagonal basis $\{D,A\}$.
For this particular measurement, indium-gallium-arsenide (InGaAs) single photon detectors are employed.
Tuning of $\phi$ is achieved by changing the optical path length of the long arm of the transcriber using the PZT.
As shown in \figurename{~\ref{fig5}}, we obtain, as a function of $\phi$, interference patterns in the net coincidence rates for the three employed filters.
Here, 'net' indicates that accidental coincidence events associated with the dark counts in the InGaAs detectors have been subtracted. 

The related visibilities are 99.9$\pm$1.2, 99.4$\pm$1.5, an 97$\pm$2\% for the 100\,GHz, 540\,MHz, and 25\,MHz, respectively. The corresponding raw visibilities are 99.9$\pm$1.2, 97.1$\pm$1.2, and 88$\pm$2\%. Note that the latter raw value is only limited by the noise in the employed InGaAs detectors (see \Sectionname{~\ref{Sec_Ann_SPD}}).
These pertinent results underline the high phase stability ($\Delta \phi < \frac{\pi}{100}$) achieved with the transcriber apparatus, even for long-term measurements.\\
\indent
Furthermore, as outlined in \Sectionname{~\ref{Sec_Ann_Trans}}, the achieved accurate phase control enables switching from the $\ket{\Phi^{+}}$ ($\phi=0$) to the $\ket{\Phi^{-}}$ ($\phi=\pi$) Bell states (see \figurename{~\ref{fig5}}), or creating any superpositions of these two states, with switching speeds of more than 1\,kHz. \\
\indent
Eventually note that such a high stability would permit increasing the PM fibre length from 18\,m to $\sim$\,90\,m, and, therefore, manipulating single-photons down to $\sim 5\rm \,MHz$ of spectral bandwidth while maintaining high quality entanglement.

\subsection{Phase-shifted fibre Bragg grating filters}
\label{Sec_Ann_FBG}

Phase-shifted fibre Bragg gratings (PS-FBG) are fibre equivalents to bulk optical cavities. Usually, they are fabricated by inserting a $\pi$ phase-shift defect in the middle of a fibre Bragg mirror. Compared to bulk optical cavities, PS-FBG filters are easy to implement since both frequency stability ($\lesssim\,$1\,MHz) and accurate tunability ($\Delta \nu / \Delta T \approx$\,200\,\rm MHz/K) are achieved using basic temperature control. However, narrowband PS-FBGs cannot directly be applied to polarisation entangled photons, as fibre birefringence would associate polarisation states with transmitted wavelengths and reduce, as a consequence, entanglement purity. To avoid this effect, we place the filtering stage right after the PPLN/W, \textit{i.e.} where both photons have the same polarisation state ($\ket V_1 \ket V_2$). Using a fibre polarisation controller (PC$_1$), the bi-photon state is oriented along one of the filter's fibre axis, such that the two photons experience the same filtering behaviour.\\
In our experiment, we utilize two different PS-FBGs centred at 1560\,nm, one featuring 58\% transmission and 540\,MHz bandwidth (AOS GmbH), and the other, 72\% transmission and 25\,MHz  bandwidth (Teraxion).\\
\indent
In the case of non-degenerate photon-pair emission, a filtering stage involving two PS-FBGs, arranged in a dual channel configuration thanks to two standard WDMs, would be necessary.
Note that no further phase stabilization would be required for this arrangement, since it would be placed before the paired photons are subjected to the transcriber.

\subsection{Employed single-photon detectors (SPD)}
\label{Sec_Ann_SPD}

For the measurements displayed in Figures~\ref{fig3} and \ref{fig5}, Alice and Bob employ each a free-running indium-gallium-arsenide (InGaAs, IDQ-220) avalanche photodiode as single-photon detector (SPD). Each detector features 20\% detection efficiency and 10$^{-6}$/ns probability of dark-count. Such a level of dark-count is reasonably low for recording the data of the experiments associated with Figures~\ref{fig3} and \ref{fig5}, which are considered as preliminary characterizations. They are also suitable for measuring the quality of the produced entanglement when considering single-photon bandwidths of 100\,GHz and 540\,MHz, that require relatively short integration times for entanglement post-selection as the coincidence peaks are narrow (see Figure~\ref{fig3}). However, utilizing the 25\,MHz bandwidth filter requires considerably longer integration times, thus increasing the probability of detecting a dark count. This strongly reduces the signal to noise ratio. To circumvent this problem for the entanglement measurement displayed in \figurename{~\ref{fig4}}, the two InGaAs SPDs are advantageously replaced by two super-conducting devices (Scontel TCOPRS-001). Compared to InGaAs SPDs, their figures of merit are a reduced detection efficiency of 7\% but a much lower dark-count probability of 10$^{-8}$/ns.
A review article on state-of-the-art SPD techniques in the framework of optical quantum information applications, reporting advantages and disadvantages of each system depending on the operation wavelength, has recently been published~\cite{Hadfield_SPD_2009}.

%

\end{document}